\documentclass[12pt,twoside]{article}
\usepackage[mathscr]{eucal}
\usepackage{amsmath,amsfonts,amssymb,amsthm,mathabx,empheq}
\usepackage{times}
\usepackage{pdfsync}
\usepackage{cite}
\usepackage{url}
\usepackage{hyperref}
\usepackage{tensor}
\usepackage{color}
\usepackage{multicol}
\usepackage{bbold}

\advance\textheight\topskip
\allowdisplaybreaks[1]

\newcommand\td{\text{d}}
\newcommand\cO{{\cal O}}
\newcommand{\p}{\partial}

\newcommand{\be}{\begin{equation}}
\newcommand{\ee}{\end{equation}}
\newcommand{\bea}{\begin{eqnarray}}
\newcommand{\eea}{\end{eqnarray}}

\newcommand{\nn}{\nonumber}
\newcommand*\xbar[1]{%
  \hbox{%
    \vbox{%
      \hrule height 0.5pt 
      \kern0.3ex
      \hbox{%
        \kern-0.0em
        \ensuremath{#1}%
        \kern-0.0em
      }%
    }%
  }%
}

\allowdisplaybreaks[1]

\DeclareFontFamily{OT1}{rsfs}{} \DeclareFontShape{OT1}{rsfs}{m}{n}{
<-7> rsfs5 <7-10> rsfs7 <10-> rsfs10}{}
\DeclareMathAlphabet{\mycal}{OT1}{rsfs}{m}{n}

\hypersetup{
colorlinks=true,
linktoc=page,    
linkcolor=blue,
citecolor=blue,
urlcolor=blue,
colorlinks=true,
}

\begin{document}

\title{Note on post-Minkowskian expansion and Bondi coordinates}

\author{Pujian Mao and Baijun Zeng}
\date{}

\def\mytitle{Note on post-Minkowskian expansion and Bondi coordinates}

\addtolength{\headsep}{4pt}

\begin{centering}

  \vspace{1cm}

  \textbf{\large{\mytitle}}

  \vspace{1cm}

  {\large Pujian Mao and Baijun Zeng}

\vspace{0.5cm}

\begin{minipage}{.9\textwidth}\small \it  \begin{center}
     Center for Joint Quantum Studies, Department of Physics,\\
     School of Science, Tianjin University, 135 Yaguan Road, Tianjin 300350, China
 \end{center}
\end{minipage}

\end{centering}

\begin{center}
Emails:  pjmao@tju.edu.cn,\,zeng@tju.edu.cn
\end{center}

\begin{center}
\begin{minipage}{.9\textwidth}
\textsc{Abstract}: In this note, we transform the linear order (at order $G$) metric from a system of pointlike bodies source in the post-Minkowskian expansion to the Bondi coordinates. We show that the Bondi 4-momentum and angular momentum coincide with the relativistic definitions of 4-momentum and angular momentum for the system of pointlike bodies. The angular momentum computed at the null infinity is free of supertranslation ambiguity.

\end{minipage}
\end{center}

\thispagestyle{empty}


\section{Introduction}

The direct observation of gravitational waves by the LIGO and Virgo collaborations \cite{LIGOScientific:2016aoc} opens new avenues for gravitational-wave astronomy. The increasing sensitivity of gravitational wave detection challenges the accurate theoretical modeling of gravitational collisions  \cite{Buonanno:2022pgc}, such as the post-Minkowskian (PM) expansion based on successive approximations in powers of the Newton constant $G$. Dissipation is the key feature of the system with gravitational radiation where the radiated 4-momentum and angular momentum are very important ingredients, see, e.g., recent investigations \cite{Damour:2011fu,Blanchet:2013haa,Blanchet:2018yqa,Ashtekar:2019viz,Damour:2020tta,Jakobsen:2021smu,Mougiakakos:2021ckm,Herrmann:2021lqe,Herrmann:2021tct,Bini:2021gat,Riva:2021vnj,Veneziano:2022zwh,Manohar:2022dea,DiVecchia:2022owy,Bini:2022enm,Bini:2022wrq,Heissenberg:2022tsn,Heissenberg:2023uvo}. The PM computations of radiated 4-momentum and angular momentum are based on the assumption of balance between the radiation from the gravitational source and the radiative fluxes at the future null infinity. 

The Einstein theory can be formulated as a characteristic initial value problem \cite{Bondi:1962px,Sachs:1962wk} when applying the Bondi gauge and coordinates system at null infinity where the gravitational radiation is characterized by the news functions, and the 4-momentum and angular momentum can be defined from asymptotic analysis \cite{Bondi:1962px,Sachs:1962wk,Newman:1965ik,Newman:1968uj,Winicour,Geroch:1977big,Prior,Ashtekar:1978zz,Ashtekar:1979xeo,Ashtekar:1979iaf,Ashtekar:1981bq,Geroch:1981ut,Dray:1984rfa}, see also recent studies \cite{Barnich:2011mi,Flanagan:2015pxa}. Naturally, the radiated 4-momentum and angular momentum can be derived by translating the gravitational wave solution in the PM expansion to the Bondi coordinates and computing the 4-momentum and angular momentum flux at null infinity. Hence, the key step is the transformation from the harmonic gauge in the PM expansion to the Bondi gauge. The generic framework was recently presented in \cite{Compere:2019gft,Blanchet:2020ngx,Blanchet:2023pce} with special applications for the multipole expansion. In this note, we will follow the framework of \cite{Compere:2019gft,Blanchet:2020ngx,Blanchet:2023pce} to transform the full linear order perturbative metric of a system of pointlike bodies source to the Bondi coordinates. 

Part of the metric components was already revealed in the Bondi coordinates with very important applications in previous investigations. The asymptotic shear was derived in \cite{Veneziano:2022zwh} to resolve the angular momentum loss puzzle in the PM expansion. The Bondi mass aspect was derived in \cite{Javadinezhad:2022ldc} for constructing a supertranslation invariant definition of angular momentum loss at $\cO(G^2)$. Here, we finish the last missing piece, the angular momentum aspect which is the most complicated part. Thus, we uncover the complete linear metric in the Bondi coordinates. A direct application of our results is to compute the Bondi angular momentum at null infinity. Surprisingly, the angular momentum is free of supertranslation ambiguity \cite{Penrose,Held}, though the linear metric has clear supertranslation arbitrariness in the Bondi coordinates \cite{Veneziano:2022zwh}. The Bondi angular momentum recovers precisely the Arnowitt–Deser–Misner (ADM) angular momentum which is noting but the relativistic definition of the angular momentum for the system of pointlike bodies. In the full Einstein theory, the coincidence of the Bondi and ADM angular momentum can be only in the canonical gauge where the asymptotic shear tensor vanishes at the far past of the null infinity
\cite{Ashtekar:1979iaf,Ashtekar:1979xeo,Ashtekar:2023wfn,Ashtekar:2023zul}, see also \cite{Veneziano:2022zwh} for discussions relevant to PM expansion. Our results indicate that the asymptotic charges in the Bondi coordinates, though defined from the order $G$ metric,  are irrelevant to order $G$ supertranslation. This reveals a very novel feature of the PM expansion for the definition of the asymptotic charges. When the supertranslation is also expanded in $G$, the supertranslation at certain order in $G$ can only affect the metric at one order higher from the charge analysis. The coincidence of the Bondi and ADM angular momentum in the PM expansion only requires that the asymptotic shear vanishes at zeroth order (at order $G^0$). Our results provide a very valuable complement to the study of matching conserved quantities at the far past of the future null infinity to the spatial infinity \cite{Ashtekar:1979iaf,Ashtekar:1979xeo,Ashtekar:1978zz,Ashtekar:2023wfn,Ashtekar:2023zul,Compere:2023qoa} from the PM perspective and could be an important stepping stone for the future investigation of relevant issues about asymptotic charges at $\cO(G^2)$.

The organization of this note is as follows. In section \ref{linear}, we review the linear metric in harmonic gauge in both Minkowski coordinates and flat-space Bondi coordinates. Some useful identities are also presented. In section \ref{Bondi}, we derive the transformation from the flat-space Bondi coordinates to the full Bondi coordinates and compute the linear metric in Bondi coordinates. In section \ref{momentum}, we derive the 4-momentum at null infinity for the linear metric which recovers precisely the ADM 4-momentum. In section \ref{AM}, we compute the angular momentum at null infinity which recovers again the ADM definition. In section \ref{II}, we comment on the effect from gravitational interaction and the supertranslation ambiguity of angular momentum. We conclude in the last section.


\section{The linear metric of pointlike bodies source}
\label{linear}

The inverse metric of the spacetime with pointlike bodies source in the harmonic gauge in the Minkowski coordinates $x^\mu=(x^0,x^i)$ is given by \cite{Damour:2020tta,Bonga:2018gzr,Veneziano:2022zwh}
\be\label{metric}
g^{\mu\nu}=\eta^{\mu\nu} - 4G \sum \frac{m}{\Gamma(x)}(v^\mu v^\nu + \frac12 \eta^{\mu\nu}),
\ee
where
\be\label{Gamma}
\Gamma(x)=\sqrt{(x^\mu-c^\mu)(x^\nu-c^\nu)\Pi_{\mu\nu}},\quad \quad \Pi_{\mu\nu}=\eta_{\mu\nu}+v_\mu v_\nu.
\ee
The sign $\sum$ denotes the sum of contributions from all particles. $v^\mu$ and $c^\mu$ are the 4-velocities and initial positions of each particle. The trajectory of the particle with respect to its proper time $\tau$ is
\be
x^\mu(\tau)=c^\mu + v^\mu \tau, \quad \quad v^2=-1.
\ee
Each particle is with energy $E$ and momentum $p^i$. The 4-momentum of every single particle is defined as $p^\mu=(E,p^i)$. The 4-momentum and angular momentum of the system are the same as the ADM definition \cite{Veneziano:2022zwh},
\be
P^\mu_{\text{ADM}}=\sum p^\mu,\quad \quad J^i_{\text{ADM}}={\epsilon^i}_{jk} \sum c^j p^k.
\ee
We will write the metric \eqref{metric} in Bondi gauge following the generic treatment for PM expansion \cite{Compere:2019gft,Blanchet:2020ngx,Blanchet:2023pce}. Here, we will not take the center of mass frame applied in \cite{Veneziano:2022zwh}. 

We first introduce the flat-space Bondi coordinates $(u,r,x^{A})$ which are connected to the Minkowski coordinates $(t,x^i)$ by
\be
t=u+r,\quad \quad x^i=r n^i(x^A),\quad \quad n^i n_i=1.
\ee
One can introduce a null 4-vector $n^{\mu}=(1,n^{i})$ pointing from $r=0$ to $x^{\mu}$.
In $(t,x^i)$ coordinates, $v_\mu=(v_t,v_i)$ is a constant vector. We list some useful formulas in flat-space Bondi coordinates $(u,r,x^A)$ for later convenience 
\be
\begin{split}
&D_A D_B (n\cdot v)=-\gamma_{AB}(v^0 + n\cdot v),\quad \quad D_A D_B (n\cdot c)=-\gamma_{AB}(c^0 + n\cdot c),\\
&D_A (n\cdot v) D^A (n\cdot v)=-\left[1+(n\cdot v)^2 + 2 v^0 (n\cdot v)\right],\\
&D_{A}(n\cdot v)D^{A}(n\cdot c)=(v\cdot c)-(n\cdot v)(n\cdot c)-v^{0}(n\cdot c)-c^{0}(n\cdot v),
\end{split}
\ee
where $\gamma_{AB}$ is the boundary metric on the celestial sphere in Bondi gauge, $D_A$ is the derivative with respect to $\gamma_{AB}$, and the dot product is defined for the summation of four components. The components of the inverse metric \eqref{metric} in flat-space background Bondi coordinates are
\be\label{flatbondi}
\begin{split}
g^{uu}&=-4G\sum{\frac{m}{\Gamma}(v_{\mu}n^{\mu})^{2}},\\
g^{ur}&=-1-4G\sum{\frac{m}{\Gamma}(-v_{\mu}n^{\mu}v_{i}n^{i}-\frac{1}{2})},\\
g^{uA}&=\frac{4G}{r}\sum{\frac{m}{\Gamma}v_{\mu}n^{\mu} D^{A}( v_{\mu}n^{\mu})},\\
g^{rr}&=1-4G\sum{\frac{m}{\Gamma}[(v_{i}n^{i})^{2}+\frac{1}{2}}],\\
g^{rA}&=-\frac{4G}{r}\sum{\frac{m}{\Gamma}v_{i}n^{i} D^{A}( v_{\mu}n^{\mu})},\\
g^{AB}&=\frac{1}{r^{2}}\gamma^{AB}-\frac{4G}{r^{2}}\sum{\frac{m}{\Gamma} D^{A}( v_{\mu}n^{\mu}) D^{B}( v_{\mu}n^{\mu})}-\frac{2G}{r^{2}}\sum{\frac{m}{\Gamma}\gamma^{AB}}.
\end{split}
\ee
Correspondingly, the metric components in the flat-space Bondi coordinates are
\be
\begin{split}
g_{uu}&=-1+4G\sum{\frac{m}{\Gamma}(v_{0}v_{0}-\frac{1}{2})},\\
g_{ur}&=-1+4G\sum{\frac{m}{\Gamma}(v_{0}v_{\mu}n^{\mu}-\frac{1}{2})},\\
g_{uA}&=4Gr\sum{\frac{m}{\Gamma}v_{0}D_{A}(v_{\mu}n^{\mu})},\\
g_{rr}&=4G\sum{\frac{m}{\Gamma}(v_{\mu}v^{\mu})^{2}},\\
g_{rA}&=4Gr\sum{\frac{m}{\Gamma}v_{\mu}n^{\mu}D_{A}(v_{\nu}n^{\nu})},\\
g_{AB}&=r^{2}\gamma_{AB}+4Gr^{2}\sum{\frac{m}{\Gamma}\left[D_{A}(v_{\mu}n^{\mu}) D_{B}(v_{\nu}n^{\nu})+\frac{1}{2}\gamma_{AB}\right]}.
\end{split}
\ee


\section{The linear metric  in Bondi gauge}
\label{Bondi}

The metric in the flat-space Bondi coordinates is not in the Bondi gauge. Now we will transform from $(u,r,x^A)$ to the coordinates $(\bar{u},\bar{r},\bar{x}^A)$ in Bondi gauge by the change of coordinates
\begin{equation}\label{coordinates}
 \bar{u}=u+{\delta}u, \quad\quad   \bar{r}=r+{\delta}r, \quad \quad  \bar{x}^A=x^{A}+{\delta}x^{A},
\end{equation}
where ${\delta}u,\,  {\delta}r,\, {\delta}x^{A}$ are $\cO(G)$. The Bondi gauge conditions, i.e., $g_{rr}=0=g_{rA}$, and determinant condition, and boundary conditions of the Bondi framework will determine ${\delta}u,\,  {\delta}r,\, {\delta}x^{A}$ from the inverse metric \eqref{flatbondi} as \cite{Veneziano:2022zwh}
\begin{align}
 \p_r \delta u&=-2G\sum{\frac{m}{\Gamma}(v_{\mu}n^{\mu})^{2}},\\  
 \p_r \delta x^A &= \frac{1}{r^2}D^A \delta u + \frac{4G}{r}\sum{\frac{m}{\Gamma}v_{\mu}n^{\mu} D^{A}( v_{\mu}n^{\mu})},\\
 -2 \delta r & = r D_A \delta x^A -2Gr \sum{\frac{m}{\Gamma} \left[ \gamma_{AB} D^{A}( v_{\mu}n^{\mu}) D^{B}( v_{\mu}n^{\mu})+1\right]}.
\end{align}
One can obtain ${\delta}u,\,  {\delta}r,\, {\delta}x^{A}$ from previous equations up to integration constants of $r$ in series expansion near null infinity. We compute up to the orders which can completely fix the complete asymptotic data at null infinity in Bondi gauge. The solutions are
\begin{align}
 {\delta}u=&2G\sum{m (n^{\mu}v_{\mu} ) \ln( r) }+\beta(u,x^{A})\nonumber\\
 &-\frac{2G}{r}\sum{m\bigg[\frac{u(1+v^{0}n^{\mu}v_{\mu})}{n^{\mu}v_{\mu}}+\frac{(c^{\mu}v_{\mu})(n^{\nu}v_{\nu})+n^{\mu}c_{\mu}}{n^{\mu}v_{\mu}}\bigg] }\nonumber\\
 &-\frac{2G}{r^2}\sum{m\frac{u^2}{4(n^{\mu}v_{\mu})^3} \left[ 3+(n^{\mu}v_{\mu})^2+6v^{0}n^{\mu}v_{\mu}+2(v^{0})^{2}(n^{\mu}v_{\mu})^{2} \right]}\nonumber\\
 &-\frac{2G}{r^2}\sum m\frac{u}{2(n^{\mu}v_{\mu})^3}\bigg[ 3n^{\mu}c_{\mu}+3c^{\mu}v_{\mu}n^{\nu}v_{\nu}-c^{0}(n^{\mu}v_{\mu})^{2}+3v^{0}n^{\mu}v_{\mu}n^{\nu}c_{\nu}\\
 &\hspace{1cm} + 2v^{0}(n^{\mu}v_{\mu})^{2}c^{\nu}v_{\nu} \bigg]  -\frac{2G}{r^2}\sum m\frac{1}{4(n^{\mu}v_{\mu})^3}\bigg[ 3(n^{\mu}c_{\mu})^{2} \nn \\
 &\hspace{1.5cm} +6n^{\mu}v_{\mu}n^{\nu}c_{\nu}c^{\sigma}v_{\sigma} -c^{\mu}c_{\mu}(n^{\nu}v_{\nu})^{2}+2(n^{\mu}v_{\mu})^{2}(c^{\nu}v_{\nu})^2 \bigg]  +\cO(\frac{1}{r^3}),\nn\\
 &\nn\\
 {\delta}x^{A}=&-\frac{2G \ln(r)}{r}\sum{m D^{A}(n^{\mu}v_{\mu})}+\alpha^{A}(u,x^{B})-\frac{1}{r}D^{A}\beta
  +\frac{2G }{r}\sum{m D^{A}(n^{\mu}v_{\mu})} \nonumber\\
 &+\frac{G}{r^{2}}\sum{m\left[\frac{1}{n^{\mu}v_{\mu}}D^{A}(n^{\nu}c_{\nu})+\frac{n^{\mu}c_{\mu} + 2n^{\mu}v_{\mu}v^{\nu}c_{\nu}}{(n^{\mu}v_{\mu})^2}D^{A}(n^{\sigma}v_{\sigma}) \right]}\nonumber \\
 &+\frac{G u}{r^{2}}\sum{m\frac{ (1+2v^{0}n^{\mu}v_{\mu} )}{(n^{\mu}v_{\mu})^2}D^{A}(n^{\nu}v_{\nu})}\nonumber\\
 &+\frac{Gu^2}{r^3}\sum{m\frac{D^{A}(n^{\nu}v_{\nu})}{2(n^{\mu}v_{\mu})^4}\left[ 1+4v^{0}n^{\mu}v_{\mu} +(n^{\mu}v_{\mu})^2+2(v^{0}n^{\mu}v_{\mu})^2 \right]} \nonumber\\
 &+\frac{Gu}{r^{3}}\sum m\frac{D^{A}(n^{\sigma}v_{\sigma})}{(n^{\mu}v_{\mu})^4}\bigg[n^{\mu}c_{\mu}+2v^{0}n^{\mu}v_{\mu}n^{\nu}c_{\nu}+2v^{\mu}c_{\mu}n^{\nu}v_{\nu} \\
 &\hspace{1cm} -c^{0}(n^{\mu}v_{\mu})^2+2v^{0}v^{\mu}c_{\mu}(n^{\nu}v_{\nu})^2 \bigg] +\frac{G}{r^{3}}\sum m\frac{D^{A}(n^{\tau}v_{\tau})}{2(n^{\mu}v_{\mu})^4}\bigg[(n^{\mu}c_{\mu})^2 \nn\\
 &\hspace{1.5cm} +4v^{\mu}c_{\mu}n^{\nu}c_{\nu}n^{\sigma}v_{\sigma}-c^{\mu}c_{\mu}(n^{\nu}v_{\nu})^2 +2(v^{\mu}c_{\mu}n^{\nu}v_{\nu})^2 \bigg]\nonumber\\
 &+\frac{G}{r^{3}}\sum{m\frac{D^{A}(n^{\sigma}c_{\sigma})}{(n^{\mu}v_{\mu})^3}\bigg[u(1+v^{0}n^{\mu}v_{\mu})+n^{\mu}c_{\mu}+v^{\mu}c_{\mu}n^{\nu}v_{\nu} \bigg]} + \cO(\frac{1}{r^4}),\nn \\
 &\nn\\
 {\delta}r=&G\sum{m \ln(r)(-2v^{0}-2n^{\mu}v_{\mu})}+G\sum{m(4v^{0}+3n^{\mu}v_{\mu})} +\frac{1}{2}D^{2}\beta - \frac{r}{2} D_{A}\alpha^{A} \nonumber\\
  &+\frac{Gu}{r}\sum{m\frac{1}{(n^{\mu}v_{\mu})^{3}}\left[-1+2v^{0}(n^{\mu}v_{\mu})^{3}+(n^{\mu}v_{\mu})^{2}+2(v^{0}n^{\mu}v_{\mu})^{2} \right]}\nonumber\\
  &+\frac{G}{r}\sum{m\frac{n^{\nu}c_{\nu}}{(n^{\mu}v_{\mu})^{3}}\left[ -n^\mu c_\mu +c^\mu v_\mu n^{\nu}v_{\nu} \right]} \\
  &+\frac{G}{r}\sum m\frac{1}{(n^{\mu}v_{\mu})^{2}}\bigg[ c^{0}n^{\mu}v_{\mu}+2(n^{\mu}v_{\mu})^{2}v^{\nu}c_{\nu}+2v^{0}n^{\mu}v_{\mu}v^{\nu}c_{\nu}\nn \\
  & \hspace{2cm} +2n^\mu c_\mu n^\nu v_\nu + v^0 n^{\mu}c_{\mu} \bigg] +\cO(\frac{1}{r^{2}}),\nn
\end{align}
where we apply the asymptotic form of $\Gamma(x^\mu)$ obtained from inserting the Bondi coordinates into \eqref{Gamma} near the null infinity
\begin{equation}
 \frac{1}{\Gamma}=-\frac{1}{n^{\mu}v_{\mu}r}-\frac{K(u,x^{A})}{r^{2}}-\frac{W(u,x^{A})}{r^{3}}+\cO(\frac{1}{r^{4}}),
\end{equation}
where
\begin{align}
 K(u,x^{A})&=\frac{u(1+v^{0}n^{\mu}v_{\mu})}{(n^{\mu}v_{\mu})^{3}}+\frac{(c^{\mu}v_{\mu})(n^{\nu}v_{\nu})+n^{\mu}c_{\mu}}{(n^{\mu}v_{\mu})^{3}},\\
 W(u,x^{A})&=\frac{u^2[3+(n^{\mu}v_{\mu})^2+6v^{0}n^{\mu}v_{\mu}+2(v^{0}n^{\mu}v_{\mu})^{2}]}{2(n^{\mu}v_{\mu})^{5}}\nonumber\\
 &+\frac{u[3n^{\mu}c_{\mu}+3c^{\mu}v_{\mu}n^{\nu}v_{\nu}-c^{0}(n^{\mu}v_{\mu})^{2}+3v^{0}n^{\mu}v_{\mu}n^{\nu}c_{\nu}+2v^{0}(n^{\mu}v_{\mu})^{2}c^{\nu}v_{\nu}]}{(n^{\mu}v_{\mu})^{5}}\nonumber\\
&+\frac{3(n^{\mu}c_{\mu})^{2}+6n^{\mu}v_{\mu}n^{\nu}c_{\nu}c^{\sigma}v_{\sigma}-c^{\mu}c_{\mu}(n^{\nu}v_{\nu})^{2}+2(n^{\mu}v_{\mu})^{2}(c^{\nu}v_{\nu})^2}{2(n^{\mu}v_{\mu})^{5}}.
\end{align}
The fall-off conditions from of the Bondi gauge are
\be
\bar{g}^{\bar r \bar A}=\cO(\bar {r}^{-2}),\quad\quad \bar{g}^{\bar A \bar B}=\frac{\bar\gamma^{\bar A \bar B}}{\bar{r}^2} + \cO(\bar {r}^{-3}),\quad\quad \bar{g}^{\bar u \bar r}=-1+\cO(\bar {r}^{-2}).
\ee
The first condition yields that $\p_u\alpha^A=0$. The second one determines that $\alpha^A$ is a conformal Killing vector of the celestial sphere. From global consideration, $\alpha^A$ can only be isomorphic to the Lorentz transformation. Since we assume that $\alpha^A$ is $\cO(G)$, they are the gauge transformation of linear metric induced from the diffeomorphism of the full theory during the linearization and are irrelevant to the change of inertial frame in the bulk of the spacetime. The last fall-off condition finally fixes $\beta$ as $\beta=\bar \beta(x^A)+\frac{u}{2} D_A \alpha^A$. Hence, $\bar \beta$ represents the supertranslation at the linear order. $\bar \beta$ and $\alpha^A$ characterize the BMS transformation \cite{Bondi:1962px,Sachs:1962wk,Sachs:1962zza} at the linear level.

Now we are ready to derive the metric components in the Bondi coordinates $(\bar{u},\bar{r},\bar{x}^A)$. The ${\bar g}_{\bar A \bar B}$ components were derived in \cite{Veneziano:2022zwh}. The ${\bar g}_{\bar u \bar u}$ component was revealed in \cite{Javadinezhad:2022ldc}. Now we will complete the whole task with the remaining components. In particular, the $\bar{g}_{\bar{u}\bar{A}}$ component is
\begin{align}
\bar{g}_{\bar{u}\bar{A}}=&g_{\mu\nu}\frac{{\partial}x^{\mu}}{\partial \bar{u}}\frac{{\partial}x^{\nu}}{\partial \bar{x}^{A}}=g_{\mu\nu}(\delta^{\mu}_{u}-\partial_{u}{\delta}x^{\mu})(\delta^{\nu}_{A}-\partial_{A}{\delta}x^{\nu}) \nn \\
 =&g_{uA}-r^{2}\gamma_{AB}\partial_{u}{\delta}x^{B}+\partial_{A}({\delta}u+{\delta}r) \nn \\
 =&G\sum{m\left[ 3-\frac{1-2v^{0}n^{\mu}v_{\mu}}{(n^{\nu}v_{\nu})^{2}} \right]D_{A}(n^{\mu}v_{\mu})}+D_{A}\bar\beta+\frac{1}{2}D_{A}D^{2}\bar\beta+\cO(\frac{1}{\bar{r}^2}) \nn \\
 &+\frac{2}{3\bar{r}}\left\{\bar{u}G\sum{m\frac{3D_{A}(n^{\sigma}v_{\sigma})}{(n^{\nu}v_{\nu})^4}}+G\sum{m\frac{3D_{A}(n^{\sigma}c_{\sigma})}{2(n^{\mu}v_{\mu})^3}\left[-1-2n^{\mu}c_{\mu}+v^{\mu}c_{\mu}n^{\nu}v_{\nu} \right]}\right\} \nn \\
 &+\frac{2}{3\bar{r}}G\sum{m\frac{3D_{A}(n^{\sigma}v_{\sigma})}{2(n^{\nu}v_{\nu})^4}\left[-n^{\mu}c_{\mu}+3(n^{\mu}c_{\mu})^2-2v^{\mu}c_{\mu}n^{\nu}v_{\nu}-2v^{\mu}c_{\mu}n^{\nu}c_{\nu}n^{\sigma}v_{\sigma} \right]}.
\end{align}
And we have confirmed that $\bar{g}_{\bar{u}\bar{r}}= -1 +  \cO(\frac{1}{\bar{r}^{2}})$ in the above analysis for deriving the coordinates transformation. Gathering all components in the Bondi gauge, the final results can be summarized with some massages as
\begin{align}
 \bar{g}_{\bar{u}\bar{u}}=& -1 - \frac{2G}{\bar{r}}\sum{\frac{m}{(v_{\mu}n^{\mu})^{3}}}+  \cO(\frac{1}{\bar{r}^{2}}),\label{uu}\\
 \bar{g}_{\bar{u}\bar{r}}=& -1 +  \cO(\frac{1}{\bar{r}^{2}}),\\
 \bar{g}_{\bar{A}\bar{B}}= &\bar{r}^{2}\bar{\gamma}_{\bar{A}\bar{B}}-4G\bar{r}\sum{\frac{m}{n^{\mu}v_{\mu}}\left[D_{\bar{A}}(n^{\mu}v_{\mu})D_{\bar{B}}(n^{\nu}v_{\nu})-\frac{1}{2} \bar{\gamma}_{\bar{A}\bar{B}} D_{\bar{C}}(n^{\mu}v_{\mu})D^{\bar{C}}(n^{\nu}v_{\nu})\right]}\nonumber\\
 &+\bar{r}\left[ 2D_{\bar{A}}D_{\bar{B}}\bar\beta-\bar{\gamma}_{\bar{A}\bar{B}} D_{\bar{C}}D^{\bar{C}}\bar\beta \right] +  \cO(1),\\
 \bar{g}_{\bar{u}\bar{A}}=&G\sum{m\left[ 3-\frac{1-2v^{0}n^{\mu}v_{\mu}}{(n^{\nu}v_{\nu})^{2}} \right]D_{\bar{A}}(n^{\mu}v_{\mu})}+D_{\bar{A}}\bar\beta+\frac{1}{2}D_{\bar{A}}D^{2}\bar\beta\nonumber\\
 &+\frac{2}{3\bar{r}}\left[-\bar{u}D_{\bar{A}}\left(G\sum{\frac{m}{(n^{\nu}v_{\nu})^3}}\right)-G\sum{m\frac{2}{(n^{\nu}v_{\nu})^{3}}D_{\bar{A}}(n^{\mu}c_{\mu})} \right]\nonumber\\
 &+\frac{2}{3\bar{r}} D_{\bar{A}}\left\{\frac{3G}{2}\sum{m\left[-\frac{(n^{\mu}c_{\mu})^{2}}{(n^{\nu}v_{\nu})^{3}}+\frac{v^{\mu}c_{\mu}n^{\nu}c_{\nu}}{(n^{\sigma}v_{\sigma})^{2}}+\frac{v^{\mu}c_{\mu}}{(n^{\nu}v_{\nu})^{2}}+\frac{1}{3}\frac{n^{\mu}c_{\mu}}{(n^{\nu}v_{\nu})^3} \right]} \right\}\nn\\
 &+\cO(\frac{1}{\bar{r}^2}).\label{uA}
\end{align}
One can recognize that the linear metric does not involve the parameters $\alpha^A$. This just reflects the fact that $\alpha^A$ represent Lorentz transformations which are part of the isometry of the background spacetime. In the linearization, they become the reducibility parameters for the linear metric \cite{Barnich:2001jy}. Their corresponding transformations, by definition, do not change the linear metric. Similarly, the translation parts from $\bar\beta$ do not affect the linear metric either. Comparing to the solution near null infinity in Bondi gauge,
\begin{multline}
ds^2=-\left[1-\frac{2 m_b}{r}+ \cO(r^{-2})\right]\td u^2 - 2\left[1 + \cO(r^{-2})\right]\td u \td r \\
+ \bigg[D^A C_{AB}+ \frac{4}{3 r} (N_{B}+u D_{B} m_b)-\frac{1}{8r}D_B (C_{AE}C^{AE})  \\
 + \cO(r^{-2})\bigg] \td u \td x^B
 +\left[r^2 \gamma_{AB} + r C_{AB} + \cO(1)\right]\td x^A \td x^B,
\end{multline}
one can read off the traceless shear tensor \cite{Veneziano:2022zwh} as 
\begin{multline}\label{CAB}
C_{\bar{A}\bar{B}}=-4G \sum{\frac{m}{n^{\mu}v_{\mu}}\left[D_{\bar{A}}(n^{\mu}v_{\mu})D_{\bar{B}}(n^{\nu}v_{\nu})-\frac{1}{2} \bar{\gamma}_{\bar{A}\bar{B}} D_{\bar{C}}(n^{\mu}v_{\mu})D^{\bar{C}}(n^{\nu}v_{\nu})\right]} \\
 +\left[ 2D_{\bar{A}}D_{\bar{B}}\bar\beta-\bar{\gamma}_{\bar{A}\bar{B}} D_{\bar{C}}D^{\bar{C}}\bar\beta \right],
\end{multline}
from which, one can verify
\be
 \frac12 D^{\bar B}C_{\bar A \bar B}=G\sum{m\left[ 3- \frac{1 - 2v^{0}n^{\mu}v_{\mu}}{(n^{\nu}v_{\nu})^{2}} \right]D_{\bar A}(n^{\mu}v_{\mu})} + D_{\bar A}\bar\beta+ \frac12 D_{\bar A}D^{2}\bar\beta,
 \ee
and 
\be\label{identity}
D_{\bar B} D_{\bar A} D_{\bar E} C^{\bar B \bar E}-D^2 D^{\bar B} C_{\bar A \bar B}=0.
\ee
The Bondi mass aspect \cite{Javadinezhad:2022ldc} is obtained as 
\be
m_b=-G\sum{\frac{m}{(v_{\mu}n^{\mu})^{3}}}.
\ee
The angular momentum aspect is
\be
N_{\bar A}=-G\sum{m\frac{2}{(n^{\nu}v_{\nu})^{3}}D_{\bar A}(n\cdot c)}+\text{total derivative terms}.
\ee
The total derivative terms can be read from previous equations for the metric components. We omit them here because they will not contribute to the angular momentum at null infinity. It is important to notice that this expression of the angular momentum aspect is the main new result of this note. To obtain that, we have computed $\delta x^A$ with two more orders than \cite{Veneziano:2022zwh} and $\delta u$ with one more order than \cite{Javadinezhad:2022ldc}. In the following sections, we will compute the 4-momentum and angular momentum at null infinity from those asymptotic data.


\section{Four-momentum at null infinity}
\label{momentum}

The definition of Bondi 4-momentum is
\be
P^{\mu}=\frac{1}{4\pi G}\int_{S} \, m_b n^{\mu} \, \td\Omega=\frac{1}{4\pi}\sum{m\int_{S}\frac{n^{\mu}}{(v^{0}-n^{i}v_{i})^{3}} \td\Omega } \label{4-momentum} ,
\ee
where $\td \Omega$ is the integral measure on the celestial sphere. We can select the z-axis in the spherical integral along the direction of the 3-velocity $\vec{v}$. Then $v^x=0=v^y$ and $v^z=v=|\vec{v}|$ in our arrangement. 
The integral for the 4-momentum on the sphere is reduced to
\begin{equation}
    \int_{S}\frac{n^{\mu}}{(v^{0}-n^{i}{v_{i}})^{3}}d\Omega=\int^{2\pi}_{0}d\phi\int^{\pi}_{0}\frac{n^{\mu}\sin{\theta}}{(v^{0}-v\cos{\theta})^3}d\theta,
\end{equation}
Then we can integrate for different components as
\begin{align}
    &\int^{2\pi}_{0}d\phi\int^{\pi}_{0}\frac{n^{0}\sin{\theta}}{(v^{0}-v\cos{\theta})^3}d\theta=\int^{2\pi}_{0}d\phi\int^{\pi}_{0}\frac{\sin{\theta}}{(v^{0}-v\cos{\theta})^3}d\theta=4\pi v^{0},\\
    &\int^{2\pi}_{0}d\phi\int^{\pi}_{0}\frac{n^{z}\sin{\theta}}{(v^{0}-v\cos{\theta})^3}d\theta=\int^{2\pi}_{0}d\phi\int^{\pi}_{0}\frac{\cos{\theta}\sin{\theta}}{(v^{0}-v\cos{\theta})^3}d\theta=4\pi v,
\end{align}
and the other two components are zero due to their integration over a period involving $\sin{\phi}$ or $\cos{\phi}$. Note also that $v^x=0=v^y$ and $v^z=v$. So the final result is
\begin{equation}
    \int^{2\pi}_{0}d\phi\int^{\pi}_{0}\frac{n^{\mu}\sin{\theta}}{(v^{0}-v\cos{\theta})^3}d\theta=4\pi v^{\mu}.
\end{equation}
Finally, \eqref{4-momentum} is reduced to
\be
P^{\mu}=\sum{ m v^{\mu}},
\ee
which coincides precisely with the ADM 4-momentum.


\section{Angular momentum at null infinity}
\label{AM}

The definition of angular momentum at the leading order in $G$ is
\be
J^{i}=\frac{1}{8\pi G}\int_{S} \, Y^{i\bar A} N_{\bar A} \, d\Omega,
\ee
where $ Y^{i \bar A} \frac{\p}{\p x^{\bar A}}$ are three Killing vectors of the celestial sphere, which are related to the normal vector $n^i$ by
\begin{equation}
    Y^{(i)\bar{A}}=-\epsilon^{\bar A \bar B}D_{\bar B}n^{i}.
\end{equation}
Applying the identity
\be
\epsilon^{AB}D_{A}(n\cdot c)D_{B}(n^k)=\epsilon^{ijk}n_{i}c_{j},
\ee
we obtain
\begin{align}
    J^i&=-\frac{1}{4\pi}\sum{m\int_{S}\frac{\epsilon^{\bar{A}\bar{B}}D_{\bar{A}}(n\cdot c )D_{\bar{B}} n^{i}}{(v^{0}-n^{i}v_{i})^{3}}d\Omega }\nonumber\\
    &=-\frac{1}{4\pi}\sum{m\int_{S}\frac{\epsilon^{ijk}n_{j}c_{k}}{(v^{0}-n^{i}v_{i})^{3}}d\Omega }\nonumber\\
    &=\epsilon^{ijk}\sum{mc_{j}v_{k}},
\end{align}
where the integral on the celestial sphere can be calculated similarly as the 4-momentum. The final result recovers precisely the ADM angular momentum. Here we obtain it from the asymptotic point of view at null infinity.


\section{Comments on gravitational interaction and supertranslation ambiguity}
\label{II}

In the computations in the previous sections, we assumed that the pointlike bodies are moving along straight lines, namely the pointlike bodies have constant velocities. Correspondingly, there is no gravitational interaction at the leading order (at order $G^0$). Nevertheless, most of the results also hold in the scattering process. It is well known that the first nonzero radiative field appears at order $G^2$ for scattering of gravitating point particles. Hence, the momentum and angular momentum of the pointlike bodies are conserved at the leading order ($G^0$), which yields that the motions of the pointlike bodies are following straight lines at early/late times. The simplest mode of such motions is the instantaneous interaction case. Assume that the point particles are with velocities $v_\mu$ and $v'_\mu$ at $t < 0$ and $t > 0$ respectively. And the initial and final positions of the particles are $c_\mu$ and $c'_\mu$. The resulting spacetime metrics at $t < 0$ and $t > 0$ will have the same form as \eqref{metric} but with different velocity and different position parameters. The transformation putting the metric from harmonic gauge to Bondi gauge in section \ref{Bondi} can apply directly to both cases with different changes of coordinates, namely $\delta u$, $\delta r$, and $\delta x^A$ are different as they have clear velocity and position dependence. The final formulas of metric in Bondi gauge are precisely in the form \eqref{uu}-\eqref{uA} with $v_\mu$, $c_\mu$ for $t<0$ and $v'_\mu$, $c'_\mu$ for $t>0$. The momentum and angular momentum at null infinity can be obtained directly from the results in the previous two sections, which are given by $\sum m v^\mu$ and $\epsilon^{ijk}\sum{mc_{j}v_{k}}$ for early time, and $\sum m {v'}^\mu$ and $\epsilon^{ijk}\sum{mc'_{j}v'_{k}}$ for late time . This simply recovers the momentum and angular momentum conservation of the particles' motion at the leading order $G^0$ in the scattering process. Hence, the entire metric for the instantaneous interaction case can be transformed into Bondi gauge with distinct change of coordinates for early and late times. The momentum and angular momentum conservation are recovered from null infinity analysis in the Bondi gauge. Notably, the angular momentum is free of supertranslation ambiguity. 

To close this section, we will comment more on the supertranslation. The transformation in \eqref{coordinates} involves only $\cO(G)$ corrections. This will guarantee that the coordinates transformation \eqref{coordinates} will induce a gauge transformation of the linearized theory. While the zeroth order transformations have very different nature than the first order ones.\footnote{We did not include the coordinate transformation at order $G^0$. Here, we briefly comment on the transformation at order $G^0$.} They characterize the change of frame for the background Minkowski spacetime. A supertranslation of the Minkowski can be obtained in a closed form \cite{Compere:2016jwb}, see also \cite{Mao:2024pqz}. The linearized theory is defined in a covariant way in the Minkowski spacetime. Hence, a supertranslated perturbative metric will be a solution of the linearized theory in the Minkowski spacetime with a supertranslation. The leading order momentum and angular momentum will also be transformed in a covariant way. In particular, the leading order angular momentum will have zeroth order supertranslation dependence, i.e., the supertranslation ambiguity.\footnote{The supertranslation invariant definition of angular momentum is free of this ambiguity \cite{Compere:2019gft,Chen:2021szm,Compere:2021inq,Chen:2021kug,Javadinezhad:2022hhl,Chen:2022fbu,Javadinezhad:2022ldc}.} More precisely, a supertranslation that transforms one metric in Bondi gauge to a new metric in Bondi gauge is given by the following change of coordinates in asymptotic expansions,
\be
\begin{split}
&\bar u=u' + T({x'}^{{A}'}) - \frac{D^{{A}'} T D_{{A}'} T}{2r'} + \cO(\frac{1}{{r'}^2}),\\
&\bar r=r'+\frac12 D^{{A}'} D_{{A}'} T + \frac{D^{{A}'} T D_{{A}'} T}{2r'} + \cO(\frac{1}{{r'}^2}),\\
&\bar x^{{A}} = {x'}^{{A}'} - \frac{D^{{A}'} T}{r'} + \cO(\frac{1}{{r'}^2}),
\end{split}
\ee
where $(u',r',{x'}^{A'})$ are the new Bondi coordinates. Consequently, the transformation law of the physical quantities, i.e., the shear tensor, the Bondi mass and angular momentum aspects, for the zeroth order supertranslation are \cite{Chen:2023zpl,Flanagan:2023jio,Veneziano:2025ecv}
\be\label{transformed}
\begin{split}
&{\gamma'}_{A' B'}=\gamma_{A' B'},\qquad  {C'}_{{A}' {B}'}= {C}_{A' B'} ,\qquad {m'}_b=m_b,\\
& {N'}_{A'}=N_{A'} + 3 m_b D_{A'} T + \frac34 D_{B'} T \left(D_{A'} D_{C'} C^{B' C'} - D^{B'} D^{C'} C_{C' A'}\right).
\end{split}
\ee
Note that there is no radiation at order $G$, hence the news tensor $N_{A' B'}=0$. The full transformation of the shear tensor is ${C'}_{{A}' {B}'}= {C}_{A' B'} - 2 D_{{A}'}  D_{{B}'} T + \gamma_{A' B'} D^{C'}D_{C'} T$. If one expands the supertranslation $T$ in powers of $G$, the zeroth order supertranslation can not change the order $G$ shear \eqref{CAB}. It will turn on the shear tensor of the background metric $\eta_{\mu\nu}$, i.e., the order $G^0$ shear tensor which will arise the supertranslation ambiguity of angular momentum from two perspectives. On the one hand, the non-vanishing order $G^0$ shear tensor yields that the loss of angular momentum is at order $G$ according to the evolution equations, see e.g., \cite{Veneziano:2022zwh}
\be
\p_u J^i=-\frac{3}{16\pi G}{\epsilon^i}_{jk}\int (n^{[j} D_A n^{k]}) \left(\frac16 N_{BC} D^B C^{AC} - \frac12 N^{AC}D^B C_{BC}\right)\td S^2.
\ee
Note that this angular momentum loss puzzle is \textit{one order lower} (at order $G$) than the widely investigated angular momentum loss puzzle in literature (at order $G^2$). On the other hand, the angular momentum aspect in \eqref{transformed} has clear supertranslation dependence or ambiguity. There are two pieces on the right hand side of transformed angular momentum aspect. The first piece is the standard (super)translation covariant formula of the angular momentum. Interestingly, the second piece includes the possible mixed terms of the zeroth and first orders supertranslation as the shear tensor \eqref{CAB} involves the first order supertranslation $\beta$. However, the second piece vanishes when the shear can be written in the form of $C_{A' B' }=D_{A'} D_{B'} f - \frac12 \gamma_{A' B'} D^2 f$. Note that the commutator $[D^2, D_{A'}]f=D_{A'} f$ on the sphere should be applied for the verification. Hence, the first order supertranslation must be dropped out when computing the angular momentum and there is no ambiguity from the mixed zeroth and first order supertranslations. Actually, the full shear tensor \eqref{CAB} can be written in the form $C_{A' B' }=D_{A'} D_{B'} f - \frac12 \gamma_{A' B'} D^2 f$ for the case of a system of pointlike bodies source \cite{Veneziano:2022zwh}. Thus, there is no new ambiguity from supertranslation for the angular momentum in the case under consideration. The conclusion that there is no new ambiguity from supertranslation holds also for the scattering setup discussed above. However, the inclusion of zeroth supertranslation arises the issue that the angular momentum can be different at $t<0$ and $t>0$ if the observer switches to a different frame by supertranslation at $t=0$. The results in previous sections are only relevant to the first order supertranslation. Correspondingly, we have chosen the canonical gauge \cite{Veneziano:2022zwh,Veneziano:2025ecv} to fix the zeroth order supertranslation. Hence, the angular momentum in this investigation is free of supertranslation ambiguity.


\section{Conclusion}

In this note, we obtain the linear metric from a system of pointlike bodies source in the Bondi coordinates near future null infinity. The perturbative metric is at the linear order. The flat-space Bondi coordinates have already had the background Minkowski metric in the Bondi gauge. So the transformation from the flat-space Bondi coordinates to the full Bondi coordinates only involve $\cO(G)$ corrections. Hence, the Lorentz frame describing the system of pointlike bodies has never been touched. The linear metric in Bondi gauge involves a supertranslation parametrized by a free function on the celestial sphere. The Bondi mass aspect and angular momentum aspect are specified which do not have supertranslation dependence. The 4-momentum and angular momentum at null infinity are determined by the Bondi mass aspect and angular momentum aspect, respectively. The obtained conserved quantities at null infinity recover precisely the AMD 4-momentum and angular momentum at spatial infinity. The next step will be the quadratic order which will uniquely determine the radiated 4-momentum and angular momentum and bring new insights on some puzzling facts, such as the supertranslation ambiguity of the angular momentum at null infinity \cite{Penrose,Held}, see also some recent progresses on this issue \cite{Ashtekar:2019rpv,Javadinezhad:2022ldc,Fuentealba:2022xsz,Fuentealba:2023syb,Mao:2023evc,Riva:2023xxm,Javadinezhad:2023mtp,Gralla:2024wzr}.


\section*{Acknowledgments}

The authors would like to thank the anonymous referee for the suggestion on the instantaneous interaction case. This work is supported in part by the National Natural Science Foundation of China (NSFC) under Grants No. 12475059, No. 11935009, and No. 11905156.


\providecommand{\href}[2]{#2}\begingroup\raggedright\endgroup

\end{document}